\documentclass[12pt]{article}

\input epsf
\usepackage{epsfig,graphicx,epsf,hyperref}
\usepackage{amsmath,amsfonts,amssymb,amsbsy}
\hypersetup{colorlinks=true,linkcolor=blue,citecolor=blue,urlcolor=blue}

\def\i{\mbox{i}}
\def\\i{\mbox{\scriptsize{i}}}
\def\d{\mbox{d}}

\font\eulerFM = eufm10 scaled \magstep1
\def\N{\mbox{\eulerFM N}}

\title{PDF at small $x$ in the non-perturbative region}
\author{M. L. Nekrasov\\
{\small\it 
Institute for High Energy Physics, NRC ``Kurchatov
Institute'',}  \vspace*{-4\baselineskip}\\
{\small\it Protvino 142281, Russia} }
\date{} 

\begin{document}
\maketitle

\begin{abstract} 
Parton distribution function (PDF) at small $x$ in a fast-moving proton is investigated within an upgraded parton model that includes parton splitting with branching cascades and parton fusion. In the region of moderately small $x$, we obtain a power-law behavior of the parton density $x f(x)$ with an exponent proportional to the logarithm of the probability of parton splitting. Taking into account parton fusion leads to a nonlinear equation for the PDF. In the region of very small $x$, the phenomenon of saturation of the parton density is detected and a model estimate of its value in this regime is obtained. The results are compared with those obtained previously based on the analysis of equations in logarithmic approximations of perturbative QCD.
\end{abstract}

\section{Introduction}\label{sec1}

We study fast-moving protons (hadrons) on the scale of small transfers $Q^2$, comparable to their inverse transverse sizes. Essentially, we mean protons in a state of free or nearly free motion. The effect of fast motion is manifested in the fact that the ultrarelativistic proton can be represented as a cloud of a large number of weakly interacting partons \cite{Feynman}. The weakness is understood in an effective sense as a consequence of the relativistic dilation of the time of interaction of partons with each other. As a result, for finite time intervals partons do not have time to fully interact with each other, and behave as quasi-free particles. As a whole, the mentioned cloud of partons can be described by a static generalized distribution of partons by momenta and positions in a plane perpendicular to the direction of proton motion.

Unfortunately, on the scale of small transfers, the QCD-based description of the above distribution is greatly complicated due to the large value of the coupling constant and inapplicability of perturbation theory. Therefore, in this region the parton distributions are determined phenomenologically based on the most general considerations. In particular, in this way the parton distribution functions (PDF) in longitudinal momenta are determined. Namely, they are first determined based on assumptions about their shape in the region of relatively small $Q^2$. Then, at large $Q^2$, they are determined using the QCD evolution equations, DGLAP \cite{DGLAP1,DGLAP2,DGLAP3} or its generalizations, see e.g.~\cite{GLR,Mueller}. This is sufficient to describe many phenomena, but does not solve the problem from first principles. A complete solution requires to know the initial distributions at small $Q^2$. It is especially important to determine PDF at small $x$ and small $Q^2$, since in this region non-trivial phenomena may occur due to increasing parton density \cite{GLR,Mueller}.

For a consistent solution of the mentioned problem within the framework of QCD, non-perturbative research methods are required. However, in their absence, for preliminary analysis the parton model is suitable. It does not rely on knowledge of the details of parton interactions, but based on general properties can provide a qualitative description of their behavior. In any case, research in the parton model could help clarify what is a consequence of the idea of partons, and what is a more specific consequence of QCD. This aspect of the problem is additional and also of interest.

In this paper we consider a simple generalization of the classical parton model \cite{Feynman} (see also e.g.~\cite{IKL}), which takes into account the splitting and fusion of partons. Both processes are characterized by probabilities introduced as model parameters. At the same time, they describe underlying processes leading to the stationary parton distribution. An essential point of our analysis is taking into account the branching of parton splitting cascades and the interactions between different branches via parton fusions. In addition, we take into account the finiteness of the cascades due to the lower bound of the longitudinal momenta of partons. Here we follow the provision that partons cannot have a longitudinal momentum below a certain limit, since otherwise they cease to belong to the hadron, merging with vacuum fluctuations \cite{Gribov1973}. However, the splitting does not necessarily have to continue until the partons reach the minimum momentum. So the splitting chains can end before the minimum momentum is reached. Thus, we take into account all possible contributions regardless of the value of the splitting probability. In this sense, our consideration is indeed~non-perturbative. 

As a first step of the investigation, we consider, following \cite{Gribov1973}, a simplistic version of the model in which all the partons in a fast-moving proton are generated by a single parent parton, which represents the proton conditionally in the infinite past. This simplification obviously limits the possibilities of studying the system for large Feynman $x$. However, after a large number of evolutionary steps, the inaccuracies in the initial conditions should be largely leveled out. So, at a qualitative level, our model should adequately capture the main properties of parton evolution at sufficiently small $x$. This is enough for our purposes, since at this stage we are focused on qualitative research only. Notice also that at small x, according to modern conceptions, gluons dominate in parton distributions, while quark contributions are weakly represented. In this regard, we consider partons of only one type. Thus, we avoid non-principled technical complications of the model associated with taking into account different types of partons.

In the next section we consider the model without taking parton fusion into account. In section \ref{sec3} we introduce parton fusion and obtain a nonlinear equation for PDF. The consistency conditions of the model is also considered in section \ref{sec3}. The phenomenon of saturation of PDF in the region of very small $x$, and the main characteristics of parton distributions in this regime are discussed in section \ref{sec4}. In the last section, we compare our results with those obtained previously based on the analysis of equations in various logarithmic approximations of perturbative QCD (pQCD), such as equations of Balitsky-Fadin-Kuraev-Lipatov (BFKL) \cite{BFKL1,BFKL2,BFKL3}, Gribov-Levin-Ryskin (GLR)\cite{GLR,Mueller} and Balitskii-Kovchegov (BK)~\cite{BK1,BK2}.

\section{PDF due to parton splitting}\label{sec2}

The main concept of our analysis is the generation (age) of partons. We consider a generation to be partons that appear as a result of the same number of successive splittings of the initial parent parton. In particular, the initial parton with the longitudinal momentum $P$ belongs to the 0-th generation, and the two partons produced due to its splitting belong to the 1-st generation, etc. We also assume, in accordance with the argumentation \cite{Gribov2022}, that the longitudinal momenta of the parent partons in a freely moving proton are, on average, equally divided between the daughter partons. So, each parton of the $n$-th generation has a longitudinal momentum $P/2^n$, which corresponds to the fraction of the initial momentum $x_n=1/2^n$. The splitting continues until the longitudinal momenta of the partons become comparable to the momenta of the vacuum fluctuations. When this occurs, the splitting effectively stops, since the partons dropped out as a result of further splitting are immediately replaced by an equal number of partons captured from the vacuum (the balance is ensured by conservation of the longitudinal momentum of the proton). Based on this, we can estimate the maximum number of generations ${\bar n}$ using equation $P/2^{\bar n} = \mu$, where $\mu$ is the minimum longitudinal momentum of a parton (in order of magnitude it coincides with the inverse radius of the proton at rest). From here we get ${\bar n}(P) = \ln(P/\mu)/\ln2$. At $P=3$ TeV and $\mu = 300$ MeV this gives ${\bar n} \approx 13$.  

In this section we neglect the contributions of parton fusions (we take them into account in the next section) and consider the evolution of partons only due to their splitting. We assume that partons split independently of each other with probability $w$, common to all partons.\footnote{The constant probability of splitting is ensured by the limited virtuality of partons in a free or almost free moving proton.}  We emphasize that $w$ is an absolute probability, not a probability per unit of time. By absolute probability we mean the probability of splitting throughout the development of the quantum fluctuation to which the partons belong. We also note that for the development of splitting cascades $w \ge 1/2$ is necessary.

In principle, based on the above conditions, we can determine the probability distribution of occupation by daughter partons of each generation, from 0-th to ${\bar n}$-th. However, for our purposes this information would be redundant, since we are only interested in the average values of the occupation numbers established in the stationary state of the fast-moving proton. Note that the average occupation numbers are generally non-integer values. In particular, if the evolution started with a single initial parton, then the average occupation number of the 0-th generation is $1\!-\!w$. 

So, let us take advantage of the fact that the evolution proceeds through the splitting on average of $w$ part of the partons in each generation, and this leads to the appearance of a double number of partons in the next generation. Accordingly, the $1\!-\!w$ part of partons remains in the former generation without splitting, and they drop out of the subsequent evolution. This means that in $n$-th generation ($n \not= 0$) as a result of splitting of partons of the previous generation, on average, $\widetilde{N}_n = (2w)^n$ partons appear. Among them, if $n \not= {\bar n}$, $w \widetilde{N}_n$ partons continue to split, leading to the appearance of $\widetilde{N}_{n+1}\! = 2 w\widetilde{N}_n$ partons in the next generation, and ${N_n = (1\!-\!w)\widetilde{N}_n}$ partons remain in the $n$-th generation. For $n = 0$, we set $\widetilde{N}_0 = 1$. In the final ${\bar n}$-th generation, the total number of partons is, on average, $N_{\bar n}=\widetilde{N}_{\bar n}$. 

To check the above calculations, let us determine the total momentum of the system. Since each parton in the $n$-th generation has a momentum $x_n=1/2^n$, the total momentum is
\begin{equation}\label{ff1}
\sum_{n=0}^{{\bar n}-1} (1-w) \frac{(2w)^n}{2^n} + \frac{(2w)^{\bar n}}{2^{\bar n}} = 1\,,
\end{equation} 
as it should be. The total number of partons generated by the initial parent parton is, on average, 
\begin{equation}\label{ff2}
{\cal N} = (1-w) \sum_{n=0}^{{\bar n}-1}(2w)^n + (2w)^{\bar n} = 
1+w\frac{(2w)^{\bar n}-1}{2w-1} \,.
\end{equation} 
In the limiting case $w \to 1/2$, this gives ${\cal N} \to {\cal N}_{min} = 1+{\bar n}/2$, which means that the cascade develops without branching. This corresponds to a multiperipheral regime of splitting, in which the initial parent parton successively “drops” daughter partons until a minimum momentum is reached, and the dropped partons are not further split. At large $P$ this leads to ${\cal N}_{min} \sim \ln(P/\mu)$. In the case ${w \to 1}$, formula (\ref{ff2}) gives ${\cal N} \to {\cal N}_{max} = 2^{\bar n} = P/\mu$, which means complete splitting and the absence of partons in all generations except the last one. For intermediate $w$ and large $P$, formula (\ref{ff2}) leads to a power-law increase of the total number of partons, ${\cal N} \sim (P/\mu)^{\delta_{w}}$, where  
\begin{equation}\label{ff3}
\delta_{w} = \ln(2w)/\ln(2)
\end{equation} 
and $0<\delta_{w}<1$ when $1<2w<2$. 

Transition from a logarithmic to a faster power-law growth of the average multiplicity is a characteristic feature of the transition to regime with branching cascades. This property was previously noted, e.g.~in \cite{Polyakov}. However, in the case of soft inelastic collisions, the power-law growth of parton multiplicity does not generally lead to the same growth of multiplicity of the secondary hadron. The reason is that the parton clouds representing protons do not completely disintegrate when the parton fluctuations are interrupted due to collisions of slow partons. Really, the side parton branches represent local fluctuations independent of the ``root'' branches on which colliding partons are located. Therefore, after the root branchs are interrupted, the local side fluctuations recombine back into their parent partons, and only the root branches disintegrate. This means that secondary hadrons are produced in a multiperipheral process, resulting in a logarithmic growth in the multiplicity. Nevertheless, the number of slow partons ${N}_{\bar n}$ increases with the energy according to a power law. So, the proton scattering cross section should also increase according to a power law.

However, let us return to the result $N_n = (1\!-\!w) (2w)^n$, which describes the average number of partons in the $n$-th generation. In fact, it can be formulated in terms of a continuous parton density function. For this purpose we use the relation
\begin{equation}\label{ff4}
\Delta x_n = x_n \Delta{n}\,,
\end{equation} 
which follows at $\Delta{n}=1$ from the chain of equalities $\Delta x_n \equiv x_{n-1} - x_n = 1/2^{n-1} - 1/2^n = x_n \Delta{n}$. Next, we use the fact that the number of partons within the interval $\Delta x_n$ is equal to $f(x_n)\Delta x_n$, where $f(x)$ is the parton distribution function (PDF). On the other hand, this quantity is $N_n \Delta{n}$. Hence, we have
\begin{equation}\label{ff5}
x_n f(x_n) =  N_n \,.
\end{equation} 
It is important to note that the correspondence between descriptions in terms of continuous and discrete variables is correct only for sufficiently small $x$, or large $n$, since (\ref{ff4}) with $\Delta{n}=1$ only in this case ensures the smallness of the interval $\Delta x$, within which the number of partons is counted.

So, for sufficiently small $x$, but such that $x>x_{min} = 1/2^{\bar n}$, relation (\ref{ff5}) gives $x f(x) = (1-w) (2w)^n$. Taking into account $n=\ln(1/x)/\ln(2)$, this leads to
\begin{equation}\label{ff6}
x f(x) = (1-w) e^{-\ln(2w)\ln(x)/\ln(2)} = (1-w)\, x^{ -\delta_w }.
\end{equation} 
Recall that here the exponent satisfies $-1\!<\!-\delta_{w}\!<\!0$. In the limiting case $w \to 1/2$, formula (\ref{ff6}) yields $f(x) \sim 1/x$. In the case $w \to 1$, function $x f(x)$ describes the situation when there are no partons left at $x>x_{min}$.

Determining the realistic value of $w$ is outside the scope of our research. However, for illustrative purposes, we can estimate $w$ by relating it to specific values of $\delta_w$. Namely, we will use the widespread opinion that the exponent in the gluon PDF at $x \to 0$ is determined by the Pomeron intercept \cite{Roberts}.\footnote{This correspondence is valid in the energy region where one-pomeron exchange makes the leading contribution to the amplitude of elastic proton scattering.} Then, assuming $\delta_w = 0.1$ and $\delta_w = 0.3$, which corresponds to the so-called soft and BFKL pomeron, we obtain $w=0.54$ and $w=0.62$, respectively. With these values of $w$, the average number of partons in the system at $P=3$ TeV (which entails ${\bar n} \approx 13$) is approximately ${\cal N} \approx 12$ and ${\cal N} \approx 41$, respectively. These values, in turn, can be compared with the minimum and maximum numbers of partons in the system. At $P=3$ TeV they are ${\cal N}_{min} \approx 8$ and ${\cal N}_{max} \approx 10^4$, respectively. In the case of a soft pomeron the closeness of the average and minimum parton numbers may indicate a small contribution from cascade branching at this energy.

Concluding this section, we note that in the realistic case the number of parent partons in proton at rest is equal to three. Accordingly, the above numbers of partons must be tripled, as must the momentum for the proton. In some applications, the approximate number of partons detectable at low transfers is of great importance, see e.g.~a recent study of Coulomb contributions to the amplitude of elastic proton scattering in the framework of the Glauber approach \cite{Nekrasov2025}.

\section{Parton fusion}\label{sec3}

Let us now take into account that any two partons belonging to the same generation, i.e.~having the same longitudinal momenta, can fuse with some probability $v$, and form one new parton.\footnote{The probability of fusion of partons of different generations is extremely small due to the large difference in their momenta.} After the fusion the resulting parton, since its longitudinal momentum is doubled, passes into the previous generation, and partons that produced it disappear in the current generation. So, in each intermediate generation, partons appear both as a result of the splitting of partons of the previous generation and as a result of the fusion of partons in the next generation. Accordingly, the average number of partons that appeared in an intermediate $n$-th generation ($1 \le n \le {\bar n}-1$) is equal to
\begin{equation}\label{ff7}
\widetilde{N}_n = 2 w \widetilde{N}_{n-1} 
+ \N(v,\widetilde{N}_{n+1})\,,
\end{equation} 
where  
\begin{equation}\label{ff8}
\N(v,N) = v C_{N} + v^2 C_{N-2} + \cdots .
\end{equation}  
In (\ref{ff7}) the first term on the r.h.s.~reproduces the result of the previous section, obtained at neglecting the parton fusions. The $\N(v,\widetilde{N}_{n+1})$ is the average number of partons that appear in the $n$-th generation due to the fusion of partons of the $(n+1)$-th generation. $C_{N} = N(N\!-\!1)/2$ are binomial coefficients. The appearance of the factor $v^2$ in the second term in (\ref{ff8}) is due to the fact that the fusion of the second pair of partons occurs under the condition of the fusion of the first pair. The last term under the ellipsis corresponds to the last possible contribution with a positive binomial coefficient. In compact form, formula (\ref{ff8}) is written as follows:
\begin{equation}\label{ff9}
\N(v,N) = v \sum_{k=0}^{[N/2]-1} v^k C_{N-2k} \,,
\end{equation} 
where $[N/2]$ is the integer part of $N/2$.

In the zero and last generations, the appropriate numbers of incoming partons are 
\begin{eqnarray}\label{ff10}
& \widetilde{N}_0 = 1 + \N(v,\widetilde{N}_{1})\,,&
\\[0.5\baselineskip] \label{ff11}
& \widetilde{N}_{\bar n} = 2 w \widetilde{N}_{{\bar n}-1}\,.&
\end{eqnarray}
The first term in (\ref{ff10}) matches the initial parent parton. Condition (\ref{ff11}) implies that in the ${\bar n}$-th generation the split partons are immediately replaced by an equal number of partons from the vacuum. Note that $\widetilde{N}_{n}$ increases monotonically with $n$ at $2w\!>\!1$. 

In turn, among the $\widetilde{N}_{n}$ partons that appeared in $n$-th generation, not all remain in it. Namely, $w$ of them split and $2\N(v,\widetilde{N}_{n})$ fuse pairwise. As a result, in the $n$th generation there remain $N_n$ partons, where
\begin{equation}\label{ff12}
N_n = (1-w)\widetilde{N}_n - 2\N(v,\widetilde{N}_{n}) .
\end{equation} 
In the 0-th and last generations, the corresponding occupation numbers are 
\begin{eqnarray}\label{ff13}
N_0 \!\!&=&\!\! (1 - w) {\widetilde{N}_{0}}\,,
\\[0.5\baselineskip] \label{ff14}
N_{\bar n} \!\!&=&\!\! \widetilde{N}_{\bar n} - 2\N(v,\widetilde{N}_{\bar n})\,.
\end{eqnarray} 

Formulas (\ref{ff12})--(\ref{ff14}) describe the balance between partons incoming and leaving a given generation, resulting in some number of partons remaining in the given generation. Note that the second term in (\ref{ff12}) slows down the growth of the number of the remaining partons, and with increasing $n$ this effect becomes increasingly significant. We will discuss this issue below, but for now, let us analyze the consistency of the above description.

In fact, two consistency conditions can be formulated. The first one, which is in fact trivial, is that the sum of the probabilities for each parton to split or fuse with any other parton must not exceed 1. This means that
\begin{equation}\label{ff15}
w + v (\widetilde{N}_{n}-1) \le 1\,.
\end{equation}
The difference between the r.h.s and l.h.s. in (\ref{ff15}) is the probability that the parton remains in the $n$-th generation. To analyze consequences of (\ref{ff15}), it is convenient to rewrite it in the form
\begin{equation}\label{ff16}
\widetilde{N}_{n} \le (1-w+v)/v \,.
\end{equation} 
For given $w$ and $v$, relation (\ref{ff16}) sets a limit on the maximum value of $\widetilde{N}_{n}$ and, in view of the monotonic growth of $\widetilde{N}_{n}$, on the maximum number of evolutionary steps $n$ after which further development of cascades should be impossible.

The second condition is that the number of partons $N_n$, defined by formula (\ref{ff12}), must be non-negative. Unfortunately, to resolve this condition, we need to know the solution to equation (\ref{ff7}), but we do not know it. Nevertheless, some consequence of this condition can be obtained. Namely, let us separate in formula (\ref{ff12}) the first term of the expansion of $\N(v,\widetilde{N}_{n})$ in powers of $v$ and denote the remainder as $\Delta\N$. Then (\ref{ff12}) can be written as
\begin{equation}\label{ff17}
N_n = \widetilde{N}_n \left( 1 - w + v - v \widetilde{N}_n \right) - 2\Delta\N \,.
\end{equation} 
Next, we note that the first term in (\ref{ff17}) is non-negative under condition (\ref{ff16}), while the second term is always negative. So, in general the condition $N_n \ge 0$ imposes a stronger constraint on the maximum allowed $\widetilde{N}_{n}$ and the corresponding $n$. For small $v$, however, the difference between the two constraints should not be significant since $\Delta\N = O(v^2)$. In essence, the condition $N_n \ge 0$ establishes a constraint type relationship between the maximum allowable $n$ and the model parameters $w$ and~$v$. In the next section, we discuss the possibility of resolving this constraint and its consequences.

Here we recall that the model initially allows for a situation where the development of the cascades becomes impossible. This occurs when the number of evolutionary steps reaches the value ${\bar n}$, determined by the condition $P/2^{\bar n} = \mu$, after which the splitting of the partons effectively stops. If the corresponding $\widetilde{N}_{\bar n}$ leads to a positive $N_{\bar n}$, then evolution end in this way. But if the condition $N_n\ge 0$ is violated before reaching $\bar n$, then there must be some other mechanism for interrupting the evolution, or the evolution regime must change to one that is not covered by our model.

\section{Saturation of the parton density}\label{sec4}

Let us discuss in more detail the behavior of the system as $n$ approaches the maximum allowed value, beyond which the condition of non-negativity of $N_n$ is violated. Accordingly, we assume that the mentioned value of $n$ is less than $\bar n$, upon reaching which splitting of partons becomes effectively impossible.

To begin with we note that, in accordance with the idea of the weakness of the interaction of partons in fast-moving had\-rons, the parameter $v$ should be considered very small. In this section, we significantly use of this condition. Next, we note that the maximum value of $\widetilde{N}_{n}$ allowed by the condition (\ref{ff16}) must be large, of the order of $1/v$. For this reason, the condition of smallness of $v$ cannot be used to solve equation (\ref{ff7}) for large $n$. However, all contributions to $\N(v,\widetilde{N}_{n})$ are of the same order in $\widetilde{N}_{n}$. Therefore, given the smallness of $v$, all nonlinear in $v$ contributions to $\N(v,\widetilde{N}_{n})$ may be neglected. Unfortunately, this does not lead to a sufficient simplification of equation (7). It still remains nonlinear and its solution remains unknown. Nevertheless, the analysis of the resulting equation (\ref{ff17}) without the contribution of $\Delta\N$ can be carried out without knowing the exact solution for $\widetilde{N}_{n}$.

So, we consider equation
\begin{equation}\label{ff18}
N_n = \widetilde{N}_n \left(1 - w + v - v \widetilde{N}_n \right).
\end{equation} 
It is immediately clear from it that $N_n$, unlike $\widetilde{N}_n$, is not a monotone function of $n$, since as a function $\widetilde{N}_n$ it is described by a parabola. Accordingly, $N_n$ first grows proportionally to $\widetilde{N}_n$, and then after passing the value 
\begin{equation}\label{ff19}
\widetilde{N}_{n_{s}} = \frac{1 - w + v}{2v}, 
\end{equation}  
drops to zero as $\widetilde{N}_n$ approaches $2\widetilde{N}_{n_{s}}$. Here $n_{s}$ is a solution to equation (\ref{ff19}), in which $\widetilde{N}_{n}$ is a solution to equation (\ref{ff7}). Assuming that $\widetilde{N}_{n_s} \to (2w)^{n_s}$ as $v \to 0$, we obtain
\begin{equation}\label{ff20}
n_{s} \to \ln(1/v) \left[ \ln(2w) \right]^{-1} , \quad\; 
v \to 0\,.
\end{equation} 
The maximum value of $N_n$ by virtue of (\ref{ff18}) is equal to
\begin{equation}\label{ff21}
N_{n_{s}} = \frac{(1 - w + v)^2}{4v}.
\end{equation} 
So, both $N_{n_{s}}$ and $\widetilde{N}_{n_{s}}$ have the behavior $\sim 1/v$ as $v\to 0$, which indicates high parton density and great activity of the parton transfer between generations near $n_{s}$.

The fact that $N_n$ drops starting from $n = n_{s}$ means that the system reaches saturation in the number of partons, and then degrades approaching a state in which partons completely disappear. In principle, this is other (formally consistent) option for completing the evolution of the partons, an alternative to reaching $n$ the value of $\bar n$. The number of the last generation $n_f$ in this case is determined by the condition $\widetilde{N}_{n_f} = 2\widetilde{N}_{n_s}$, where $\widetilde{N}_{n_s}$ is given in (\ref{ff19}). However, the option of completing evolution due to depopulation against the background of high activity of transfer of partons between generations can hardly be considered satisfactory. It seems more reasonable to assume that upon reaching saturation at $n = n_{s}$, the regime of free splitting and fusion of partons is replaced by a regime in which collective effects of parton interactions become significant. 

Unfortunately, in the latter case our model provides no information about the behavior of the system. However, we can take into account that when approaching~$n_{s}$ from below the growth of $N_{n}$ slows down significantly before stopping completely. Given this, at least a slowdown in growth can be expected at $n \!>\! n_{s}$. The simplest~opti\-on is complete stabilization, when $N_{n}$ becomes a constant approximately equal to $N_{n_{s}}$. 

In any case, in the vicinity of $n \!=\! n_{s}$, where our model still works, a saturated parton medium is formed. Its characteristic feature is the high density of partons. In terms of continuous variables, the transition to this regime occurs near $x=x_s$, where  $x_s = 1/2^{n_s}$ or due to (\ref{ff20})~$x_s \sim v^{1/\delta_w}$. In this point, the parton density reaches the value of  $x_s f(x_s) = N_{n_{s}}$, where $N_{n_{s}}$ is given in (\ref{ff21}). An alternative way to describe the parton density is based on the rapidity distribution $\d N(y) / \d y$, where $y = y_{hadron} \!- \ln (1/x)$ is the rapidity of partons and $\d N(y)$ is the number of partons within interval $(y, \, y+\d y)$. Since $\d N(y) / \d y = x f(x)$, we have $\d N(y) / \d y = N_{n}$, where $y = y_{n}$. So, for $y \simeq y_{s}$ we obtain
\begin{equation}\label{ff22}
\frac{\d N(y)}{\d y} \simeq \frac{(1 - w + v)^2}{4v}\,.
\end{equation} 
Note that based on (22) the parameter $v$ can be estimated, provided that the value of $w$ is known.

In summary, we come to the following scenario. In the regime of free evolution of partons, the distribution function $x f(x)$ initially grows monotonically with decreasing $x$ (starting from a certain value of $x$ at which formula (\ref{ff5}) begins to work). Simultaneously, the density $\d N(y) / \d y$ grows with decreasing $y$. For moderately small $x$, until the nonlinear contribution in equation (\ref{ff7}) can be neglected, this growth has a power-law behavior described by formula (\ref{ff6}). But then, as $x$ decreases further, the growth rate slows down. When $x$ reaches $x_{s}$, the density of partons per unit rapidity reaches a critically high value, at which the splitting and fusion of partons ceases to be free. From this point on, the collective effects of parton interactions become significant, and a saturated dense parton medium is formed inside a fast-moving proton. The transition point to this regime is defined as the solution to equation (\ref{ff19}) with respect to $n_{s}$, where $\widetilde{N}_{n}$ is the solution to equation (\ref{ff7}). Parametrically, $n_{s}$ is defined in (\ref{ff20}). From the condition $n_{s} < \bar n$, we obtain an estimate $P_{s}/\mu \simeq v^{-1/\delta{w}}$ for the initial parton momentum starting from which relatively slow partons begin to form a saturated parton medium inside a fast proton. 

If we assume that partons at small $x$ are represented mainly by gluons, then we come to the idea of the formation of a dense gluon medium, the so-called saturated gluon matter. A similar conclusion was formulated earlier based on the analysis of equations in pQCD with summation of large logarithms \cite{GLR,Mueller,BK2}. In some studies, the saturated gluon medium inside high-energy hadrons and nuclei was called a ``color glass condensate'', see e.g.~\cite{CGC1} and review \cite{CGC2}. The parton distribution $x f(x)$ in this regime is characterized by slow growth, linear in $\ln(1/x)$. Recall that in our analysis the conclusion about the emergence of a saturated parton medium has been obtained outside the framework of perturbation theory.

\section{Discussion and conclusions}

Our analysis shows that the main properties of parton distributions in the region of small $x$ and small $Q^2$ can be described in the parton model using two parameters---the probability of parton splitting and the probability of their fusion. On a qualitative level, our results are generally consistent with those obtained previously based on the analysis of equations in log approximations of pQCD. This provides a basis for further development of the model in order to achieve a satisfactory quantitative description. On the other hand, our model clearly reveals the reasons and provides an explanation for the specific behavior of the parton distribution at small $x$. In this regard, it seems important to analyze whether there are differences in details between our results and those obtained earlier. The comparison is made below.

First of all, let us consider our result (\ref{ff6}) on the power-law behavior of the PDF in the region of moderately small $x$. We have obtained it in the approximation of neglecting the contributions of parton fusion. This approximation is reasonable in the region of not too small $x$, when the number of parton~genera\-tions $n \!\simeq\! \ln (1/x)$ has not yet reached a critically large value and the nonlinear contribution in equation (\ref{ff7}) for small values of $v$ may be neglected. Within log pQCD, a similar result with a power-law behavior of the parton distribution was obtained in the BFKL approach. The difference with our result is that the analog of the exponent $\delta_w$ in the BFKL approach is proportional to the ``decay probability'' of partons (more precisely, the coupling constant of the strong interaction of partons), while in our approach this exponent is proportional to the logarithm of the probability. The reason for the discrepancy is that our approach uses a different mechanism for the formation of the parton distribution. Specifically, in the BFKL approach the distribution function is determined based on the analysis of inelastic amplitudes in the multi-Regge kinematics, which correspond to one branch of splittings of the reggeized gluon in the $t$-channel. Accordingly, the emergence of a power law dependence of the total cross section with an exponent proportional to the coupling constant is essentially an artifact of the summation of the ladder contributions to the inelastic amplitude.\footnote{That is, the branching effect in the process of parton evolution is not taken into account.} In our approach, the parton distribution function is determined by directly counting the average number of partons in each generation, taking into account the branching of the parton cascades. It is also worth recalling that the parton model does not determine the scattering amplitude.  Instead, the cross section is directly determined by the incoherent summation of the elementary contributions of parton scattering.

In the regime with parton fusion enabled, our approach provides a nonlinear recurrence equation describing the evolution of the parton density with increasing number of generations. In the case of continuous variables, the role of the number of generations is played by $\ln(1/x)$. So, the corresponding equation for PDF should be an integro-differential equation with respect to the variable $Y =\ln(1/x)$. An equation of this type was indeed obtained in log pQCD \cite{BK2}. Moreover, the cited work also shows that this equation can be reduced to the GLR equation \cite{GLR,Mueller}, i.e. to the differential equation with respect to $\ln Q^2$ and $\ln (1/x)$. Unfortunately, analytical solutions of the above equations were obtained neither in our approach nor in log pQCD, so it is impossible to make direct comparison. However, both approaches predict the phenomenon of parton density saturation. The difference arises in the explanations of its origin. Specifically, in the pQCD approach, the occurrence of saturation is related to the scale of virtuality. Namely, for a fixed $x$ which determines the parton density, saturation occurs at such $Q^2$ when the sum of the typical parton cross sections $\sim\!\alpha_{s}(Q^2) / Q^2$ becomes comparable with the transverse area of the proton. (At the same time, it is assumed that $\alpha_{s}(Q^2)\! \ll \! 1$.) In contrast to this, in our approach, the scale of virtuality is determined by the transverse size of the proton. Therefore, the transverse sizes of partons due to the uncertainty relation are of the same order as the transverse size of the proton containing them. As a result, the overlap of the transverse area of the proton by transverse areas of partons initially occurs regardless of the value of $x$. In particular, the overlap occurs at such $x$ when there is no saturation yet. So, the fact of saturation is not directly related to the value of the virtuality and arises solely as a result of reaching a critical value of the parton density. In our approach the latter quantity is determined by the model parameters, see formulas (\ref{ff21}) and (\ref{ff22}).

Thus, there is a discrepancy between the interpretations of the occurrence of saturation in both approaches. From the point of view of our analysis, this means that either the interpretation in the framework of pQCD must be revised, or the inconsistency should be resolved when extending the interpretation to the non-perturbative region. Further research should clarify this issue.

\bigskip

\noindent {\it Acknowledgments}: The author is grateful to V.A. Petrov for valuable remarks and reference to \cite{Polyakov}.

\end{document}